\pdfoutput=1

\documentclass[preprint,showkeys,secnumarabic,amsfonts,showpacs,amsmath,amssymb]{revtex4}
\usepackage[dvips]{color}
\usepackage{epsfig}
\usepackage{amsmath}
\usepackage{graphicx}
\def\Box{\hbox{$\rlap{$\sqcup$}\sqcap$}}

\begin{document}

\title{Stability analysis and observational constraints in scalar tensor theory}
\author{Hossein Farajollahi}
\email{hosseinf@guilan.ac.ir}
\affiliation{$^1$Department of Physics, University of Guilan, Rasht, Iran}
\affiliation{$^2$ School of Physics, University of New South Wales, Sydney, NSW, 2052, Australia}

\author{Amin Salehi}
\email{a.salehi@guilan.ac.ir}
\affiliation{Department of Physics, University of Guilan, Rasht, Iran}

\author{Mohammad Nasiri}
\email{mnasiri@guilan.ac.ir}
\affiliation{Department of Physics, University of Guilan, Rasht, Iran}

\date{\today}

\begin{abstract}
We study FRW cosmology for scalar tensor theory where two scalar functions nonminimally coupled to the geometry and matter lagrangian. In a framework to study
stability and attractor solutions of the model in the phase space, we simultaneously solve the dynamical system
and best fit the stability and model parameters with the observational data. The approach imposes restrictions on the model constraints while providing information about the universe dynamics. The model predict current accelerating universe, with a phantom crossing in near future.

\end{abstract}



\maketitle

\section{Introduction}

Recently, the observations of high redshift type Ia
supernovae, the surveys of clusters of galaxies \cite{Reiss}--\cite{Riess2}, Sloan digital sky survey ({\bf
SDSS})~\cite{Abazajian} and Chandra X--ray observatory~\cite{Allen} reveal the universe accelerating expansion. Also the
observations of Cosmic Microwave Background (CMB)
anisotropies \cite{Bennett} indicate that the universe is flat and the total energy
density is very close to the critical one \cite{Spergel}. The observations though determines basic cosmological parameters
with high precisions and strongly indicates that the universe
presently is dominated by a smoothly distributed and slowly
varying dark energy (DE) component. A dynamical equation of state ( EoS) parameter that is connected directly to the evolution of the energy density in the universe and indirectly to the expansion of the Universe can be regarded as a suitable parameter to explain the universe acceleration \cite{Seljak}--\cite{Amendola}.

Motivated from string theories, the scalar-tensor models provide the simplest model-independent description of unification theories which predict couplings between scalars and curvature.
 They have assumed a prominent role since any unification scheme, such as supergravity, in the weak
energy limit, or cosmological models of inflation such as chaotic inflation, seem to be supported by them \cite{capelo}. In addition, they have been employed to study the current acceleration of the universe \cite{Sahoo}--\cite{Amendola1}.

On the other hand, by using the well-known geometric variables, Hubble parameter and deceleration parameter together with the new geometrical variables, the cosmological
diagnostic pair $\{r, s\}$ ( or statefinder parameters)\cite{Sahni}, the acceleration expansion of the universe and differentiation among the cosmological models can be explained in order to better fit the observational data. The importance of the statefinder parameters to distinct DE cosmological models is best realized, in particular, when considering the increased accuracy of the observational data during the last few years and
generality of the DE models. These parameters, in a natural next step beyond the well known geometric variables, are to differentiate the expansion dynamics
with higher derivatives of the scale factor and to explore a series of DE cosmological models,
including $\Lambda$ cold dark matter ($LCDM$), quintessence, coupled quintessence, Chaplygin gas, holographic dark energy
models, braneworld models, and so on \cite{Alam}--\cite{faraj2}. Moreover, since the cosmic acceleration affects the expansion history of the
universe, to understand the true nature of the driving force,
mapping of the cosmic expansion of the universe is very
crucial~\cite{Linder}. Hence, one requires various observational
probes in different redshift ranges to understand the expansion
history of the universe. One of these tests is the difference in distance modulus measurement of
 type Ia supernovae that helps us to testify the cosmological models.

 In this paper, we study the detailed dynamics of the nonminimally couples scalar field to gravity and matter lagrangian. Since the major difficulty in cosmological models is the nonlinearity of the field equations and thus limitation in obtaining the exact solutions, we investigate the asymptotic behaviour
of the model, which provides the relevant features to be compared with the current physical data available
for the universe. In this context, the perturbation methods, especially linear stability analysis which have been used to study the qualitative analysis of the
equations and of the long term behavior of the solutions are being proposed in this work \cite{stability}--\cite{Chiang}.

The paper organized as follows: Sec. two is devoted to a detailed formulation of the model. In Sec. three, we obtain the autonomous equations of the model and the late time attractor solutions by using the phase plane analysis in addition to best fitting the model parameters. We test the model with observational analysis; examining the behavior of the effective EoS parameter of the model and statefinders. We also reconstruct the dynamics of the coupling functions to the curvature and matter lagrangian. In Sec. four, we present summary and remarks.

\section{Nonminimally coupled scalar field model}

A general action in four dimensions where gravity and matter are nonminimally coupled to two different scalar functions given by
\begin{eqnarray}\label{ac1}
A=\int[G(\phi)R-\frac{1}{2}\phi_{;\mu}\phi^{;\mu}+L_{m}f(\phi)]\sqrt{-g}d^{4}x,
\end{eqnarray}
where the four functions $G(\phi)$ and $f(\phi)$ are not specified. Note that the four function $G$ coupled to the curvature is different from gravitation coupling constant that appears in the standard general relativity.
The action generalizes those used to construct extended inflationary models \cite{Birrel}. It is also conformally equivalent to the Brans-Dicke action in the Jordan frame by using the transformation $\varphi=G(\phi)$ and $\omega(\varphi)=\frac{G(\phi)}{2(dG(\phi)/d\phi)^2}$ and also putting $f = const$. In our units, the standard Newton coupling is also recovered in the limit $G(\phi)\rightarrow -\frac{1}{2}$ \cite{capelo}.
The field equations can be derived by varying the action with respect to $g_{\mu\nu}$
\begin{eqnarray}\label{ac2}
G(\phi)(R_{\mu\nu}-\frac{1}{2}g_{\mu\nu}R)=T_{\mu\nu}^{(\phi)}+T_{\mu\nu}^{(m)},
\end{eqnarray}
where
\begin{eqnarray}
T_{\mu\nu}^{(\phi)}=\frac{1}{2}\phi_{;\mu}\phi_{;\nu}-\frac{1}{4}g_{\mu\nu}\phi_{;\alpha}\phi^{;\alpha}-g_{\mu\nu}\Box G(\phi)+G(\phi)_{;\mu\nu}.
\end{eqnarray}
In addition, the variation with respect to $\phi$  gives the klein-Gordan equations
\begin{eqnarray}
\Box \phi+R(\frac{dG}{d\phi})+L_{m}\frac{df}{d\phi}=0
\end{eqnarray}
In FRW cosmology, the Euler-lagrange equation,corresponding to the cosmological Einstein equations are
and spatially homogeneous scalar field $\phi(t)$ are
\begin{eqnarray}
&&3H^{^{2}}=-3\frac{\dot{G}H}{G}+\frac{\dot{\phi}^{2}}{4G}+\frac{\rho_{m}f(\phi)}{2G}\label{fried1}\\
&&2\dot{H}+3H^{2}=-2\frac{\dot{G}H}{G}-\frac{\ddot{G}}{G}-\frac{\dot{\phi}^{2}}{4G}-\frac{\gamma\rho_{m}f(\phi)}{2G}.\label{fried2}
\end{eqnarray}
where eq (\ref{fried1}) is the energy constraint corresponding to the (0,0)-Enistein equation. In addition, by using the argument in \cite{mota}, for a spatially homogeneous scalar field $\phi(t)$, the field equation is
\begin{eqnarray}
&&\ddot{\phi}+3H\dot{\phi}=6(\dot{H}+2H^{2})\frac{\dot{G}}{\dot{\phi}}-(1-3\gamma)\rho_{m}f'(\phi)\label{phiequation}.
\end{eqnarray}
where prime means derivative with respect to $\phi$. In the above, we assumed that the universe is filled with the barotropic fluid
with the EoS to be $p_{m}=\gamma\rho_{m}$.
From equations (5), (6) and (7), one can easily arrive at the modified conservation equation,
\begin{eqnarray}
&&(\dot{\rho_{m}f})+3H(1+\gamma)\rho_{m}f=(1-3\gamma)\rho_{m}f'\dot{\phi}
\end{eqnarray}
In the following we assume that the matter presented in the universe is cold dark matter with $\gamma=0$.

\section{Stability analysis and best fitting}

In this section, we study the structure of the dynamical system via  phase plane analysis,
by introducing the following dimensionless variables,
\begin{eqnarray}\label{defin}
X={\frac{\dot{G}}{G H}},         Y^{2}={\frac{\dot{\phi}^{2}}{12GH^{2}}},               Z={\frac{\rho_{m}f}{6GH^{2}}}.
\end{eqnarray}
Using equations (\ref{fried1})-(\ref{phiequation}),and power law functions for $G(\phi)=\exp(\alpha\phi)$ and $f(\phi)=\exp(\beta\phi)$ the dynamical equations in terms of the new variables become,
\begin{eqnarray}
X'&=&X(\frac{\ddot{\phi}}{\dot{\phi}H}-\frac{\dot{H}}{H^{2}}),\label{x}\\
Y'&=&Y(\frac{\ddot{\phi}}{\dot{\phi}H}-\frac{X}{2}-\frac{\dot{H}}{H^{2}})\label{y}\\
Z'&=&\frac{\beta XZ}{\alpha}-3Z-XZ-2Z(\frac{\dot{H}}{H^{2}})\label{z}
\end{eqnarray}
where prime " $'$ " means derivative with respect to $N = ln (a)$ and we have,
\begin{eqnarray}
&&\frac{\dot{H}}{H^{2}}=(\frac{2Y^{2}}{4Y^{2}+X^{2}})(-3+X+X^{2}-\frac{X^{2}}{Y^{2}}-3Y^{2}+\frac{\beta ZX^{2}}{2Y^{2}}),\label{hdot}\\
&&\frac{\ddot{\phi}}{\dot{\phi}H}=-3+\frac{X}{2Y^{2}}(2+\frac{\dot{H}}{H^{2}})-\frac{\beta ZX}{2\alpha Y^{2}}\\
&&\frac{\ddot{G}}{GH^{2}}=X\frac{\ddot{\phi}}{\dot{\phi}H}+X^{2}
\end{eqnarray}
The Fridmann constraint equation (\ref{fried1}) also becomes
\begin{eqnarray}\label{constraint2}
Z=1+X-Y^{2},
\end{eqnarray}
Using the constraint (\ref{constraint2}), the equations (\ref{x})--(\ref{z}) reduce to two coupled differential equations for $X$ and $Y$. Next, we obtain the critical points (fixed points) and study
the stability of these points. Critical points are always exact constant solutions in the
context of autonomous dynamical systems. These points are often the extreme points of
the orbits and therefore describe the asymptotic behavior of the system. In the following, we find fixed points by simultaneously solving $X'=0$ and $Y'=0$. Substituting
linear perturbations $X'\rightarrow X'+\delta X'$, $Y'\rightarrow Y'+\delta Y'$ about the critical points into the two independent equations for $X$ and $Y$, to the first orders in the perturbations, yield two eigenvalues $\lambda_{i} (i=1,2)$. Stability requires the real part  of all the eigenvalues to be negative.

In the following we solve the reduced form of the coupled system of differential equations to find fixed points. From the numerical calculation we find that both critical points and eigenvalues in our model, depending on the model parameter $\alpha$ and $\beta$, are highly nonlinear. In addition, the expressions for them are long and cumbersome such that we can not evaluate under what conditions the critical points are stable or unstable. In a different approach to the problem we solve the equations by simultaneously best fitting the stability parameters and initial conditions with the observational data using the $\chi^2$ method. This affords us to find solutions for the above equations and conditions for the stability of the critical points that are physically more meaningful and observationally more favored. In the next section, we solve the equations by simultaneously best fitting the model with the observational data for distance modulus.

\subsection{Observational best fitting with distance modulus, $\mu(z)$}

The difference between the absolute and
apparent luminosity of a distance object is given by, $\mu(z) = 25 + 5\log_{10}d_L(z)$ where the Luminosity distance quantity, $d_L(z)$ is given by
\begin{equation}\label{dl}
d_{L}(z)=(1+z)\int_0^z{\frac{dz'}{H(z')}}.
 \end{equation}
 In our model, from numerical computation one can obtain $H(z)$ which can be used to evaluate $\mu(z)$. To best fit the model for the parameters $\alpha$ ,$\beta$ the initial conditions $X(0)$, $Y(0)$, $H(0)$ with the most recent observational data, the Type Ia supernovea (SNe Ia), we employe the $\chi^2$ method. We constrain the parameters including the initial conditions by minimizing the $\chi^2$ function given as
\begin{equation}\label{chi2}
 \chi^2_{SNe} ( \alpha,\beta, X(0), Y(0), H(0))=\sum_{i=1}^{557}\frac{[\mu_i^{the}(z_i|\alpha,\beta, X(0),Y(0), H(0)) - \mu_i^{obs}]^2}{\sigma_i^2},
\end{equation}
where the sum is over the SNe Ia sample. In relation (\ref{chi2}), $\mu_i^{the}$ and $\mu_i^{obs}$ are the distance modulus parameters obtained from our model and from observation, respectively, and $\sigma$ is the estimated error of the $\mu_i^{obs}$.From
numerical computation, Table I shows the best best-fitted model parameters.

\begin{table}[ht]
\caption{Best-fitted model parameters and initial conditions.} 
\centering 
\begin{tabular}{c c c c c c c } 
\hline 
parameters  &  $\alpha$  &  $\beta$ \ & $X(0)$\ & $Y(0)$\ & $H(0)$ \ & $\chi^2_{min}$\\ [2ex] 
\hline 
&$-0.23$  & $-3.2$ \ & $4.36$\ & $0.5$\ & $0.5$ \ & $545.9688383$ \\
\hline 
\end{tabular}
\label{table:1} 
\end{table}\

 The contour diagrams at the $68.3\%$, $95.4\%$ and $99.7\%$ confidence levels are given in FIG. 1). From the graph, one conclude that with $68.3\%$, $95.4\%$ and $99.7\%$ confidence level the true values for both $\alpha$ and  $\beta$ lie within the green, blue and red contours, respectively. Alternatively, we can plot the likelihood for the pair model parameters as shown in Fig. 2).

\begin{tabular*}{2.5 cm}{cc}
\includegraphics[scale=.3]{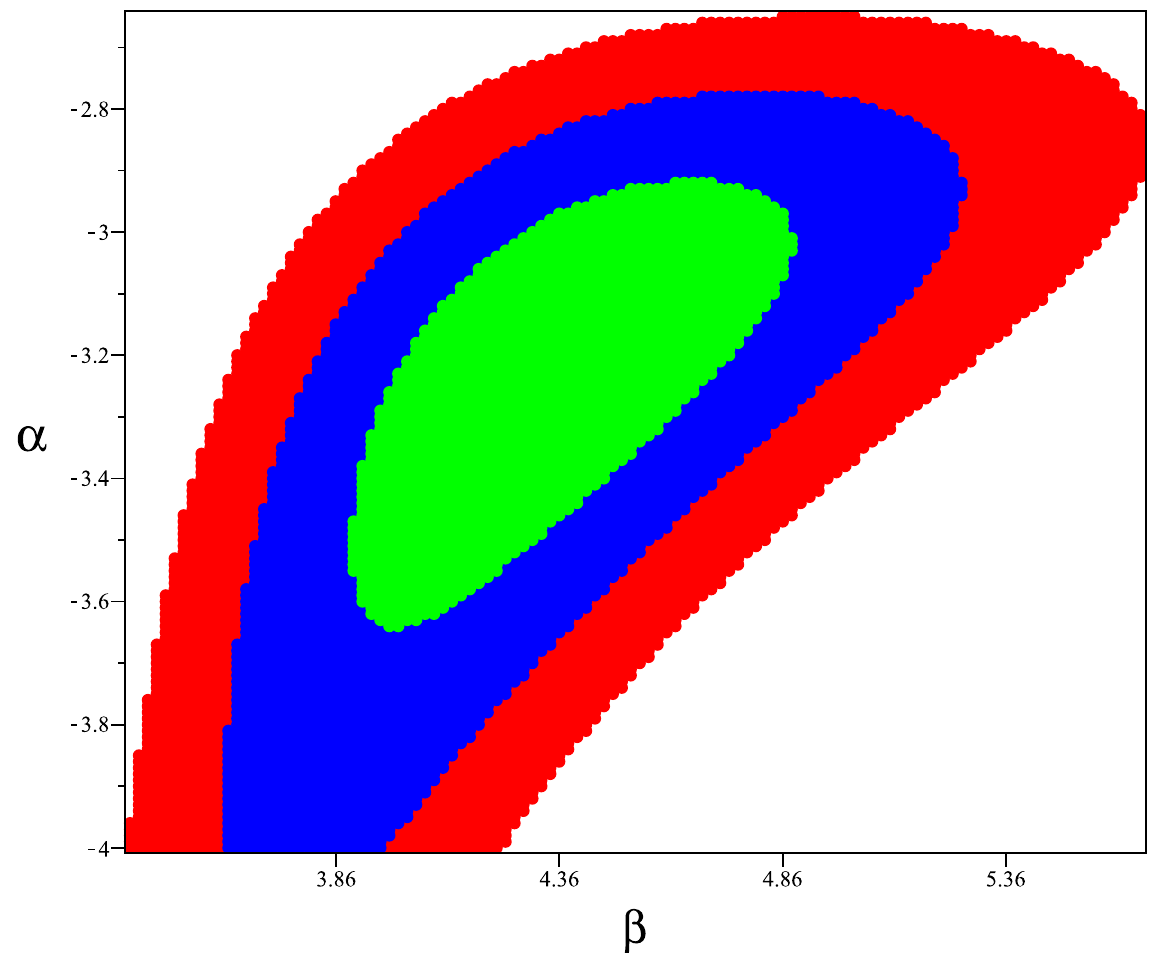}\hspace{0.1 cm}\\
Fig. 1: The constraint on $\alpha$ and $\beta$ at the 68.3\%, 95.4\% and 99.7\% confidence \\levels from Sne Ia for the model.
\end{tabular*}\\

\begin{tabular*}{2.5 cm}{cc}
\includegraphics[scale=.25]{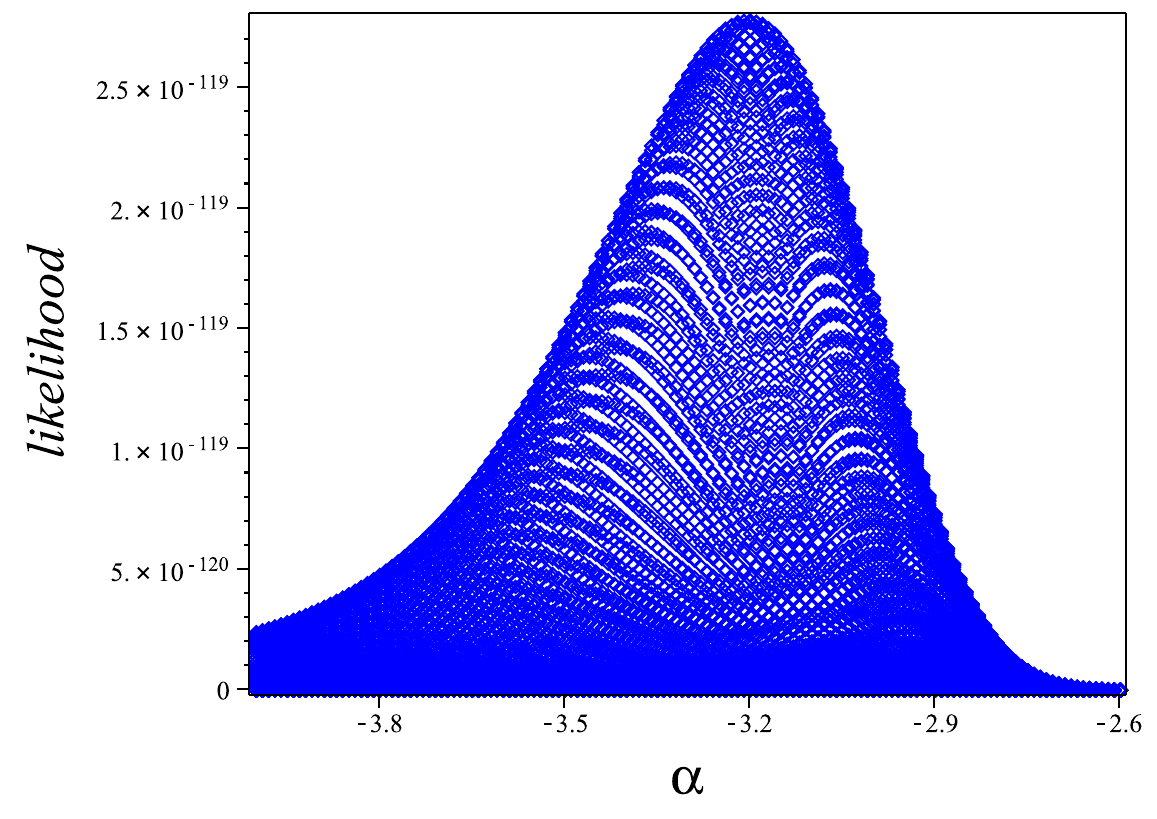}\hspace{0.1 cm}\includegraphics[scale=.25]{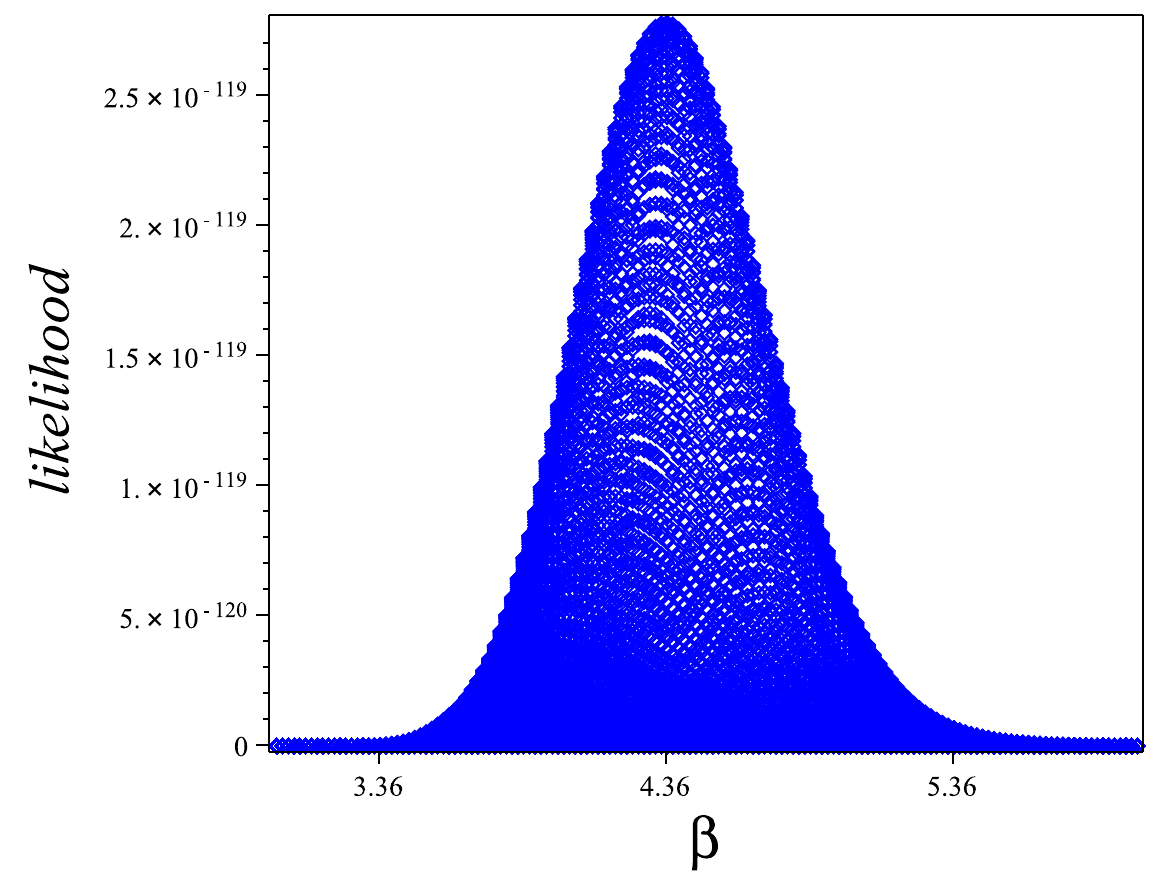}\hspace{0.1 cm}\includegraphics[scale=.3]{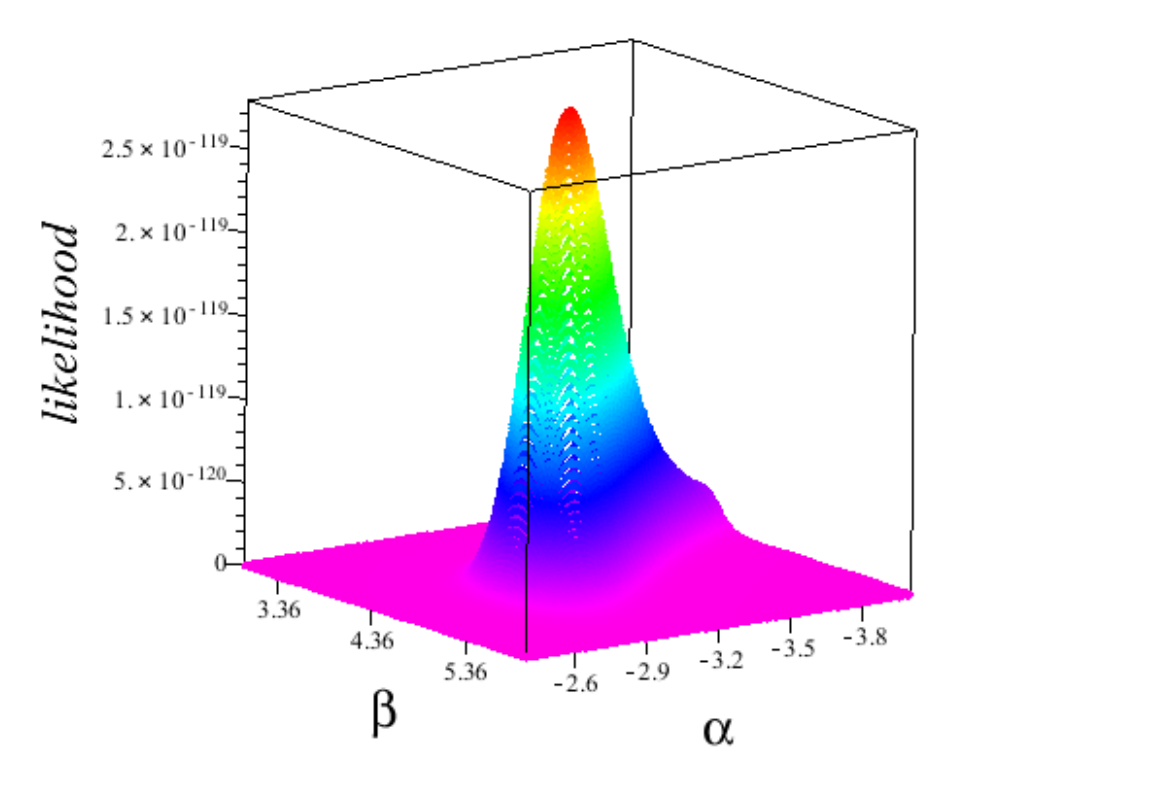}\hspace{0.1 cm}\\
Fig. 2: 1-dim and 2-dim likelihood for the model parameters
\end{tabular*}\\

In Fig. 3, the distance modulus, $\mu(z)$, in our model is compared with the observational data for the obtained parameters and initial conditions using $\chi^2$ method. As can be seen the best fitted parameters and initial conditions are in good agreement with the observational data.\\

\begin{tabular*}{2.5 cm}{cc}
\includegraphics[scale=.4]{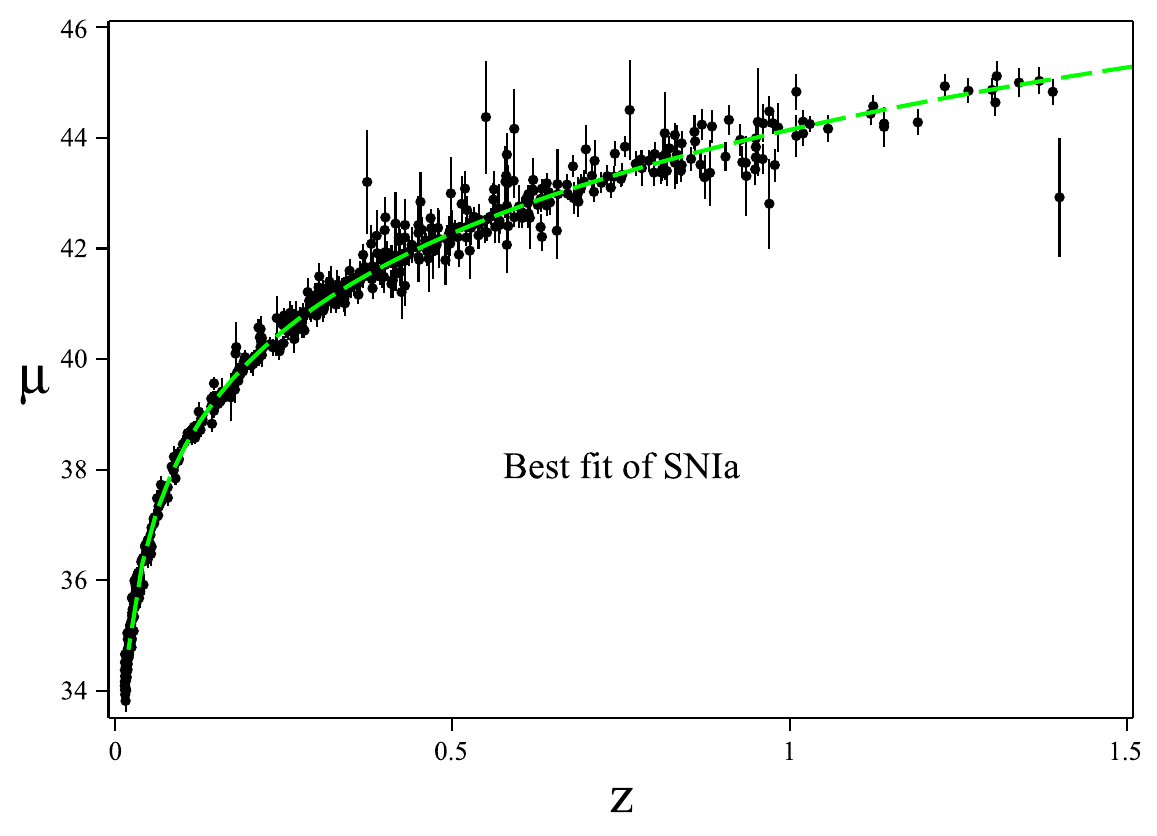}\hspace{0.1 cm}\\
Fig. 3: The best fitted distance modulus $\mu(z)$ plotted as function of redshift
\end{tabular*}\\

In the following we investigate the stability of the model
with respect to the best fitted model parameter.

\subsection{Stability of the critical points and phase space}

Solving the stability equations for the best fitted model parameter $\alpha$ and $\beta$ we find eight fixed points with the stability properties as illustrated in tables II. As can be seen the critical points in the phase space are symmetric with respect to the symmetric axis $X=0$. This is due to introduction of the dimensionless variable $Y^2$ which creates symmetric roots in the phase space. The fixed points FP7$\&$8 lie on the symmetric line, thus, is a double fixed point.

\begin{table}[ht]
\caption{est fitted critical points } 
\centering 
\begin{tabular}{c c c c c c c c c c} 
\hline\hline 
points  &  FP1  & FP2 & FP3 & FP4 & FP5 & FP6 & FP7$\&$8\\ [4ex] 
\hline 
$X  $&0.5 &0.5 &0.9 &0.9&1.5&1.5&0
 \\ 
$Y $& 0.6 & -0.6 & 1.5 & -1.5&0.9&-0.9&0
 \\
$property $  & \ \ saddle point & \ \ saddle point & \ \ unstable & \ \ unstable & \ \ stable & \ \ stable & \ \ unstable
 \\
\hline 

\end{tabular}
\label{table:1} 
\end{table}\

From the above table we see that two of the critical points, FP1 and FP2 are stable and the rest are unstable or saddle points. In Fig. 4) the trajectories leaving the unstable critical points FP5, FP6, FP7 and FP8, passing the saddle points FP1 and FP2 and finally entering the stable critical points FP5 and FP6 in the phase plane are shown.

\begin{tabular*}{3.5 cm}{cc}
\includegraphics[scale=.35]{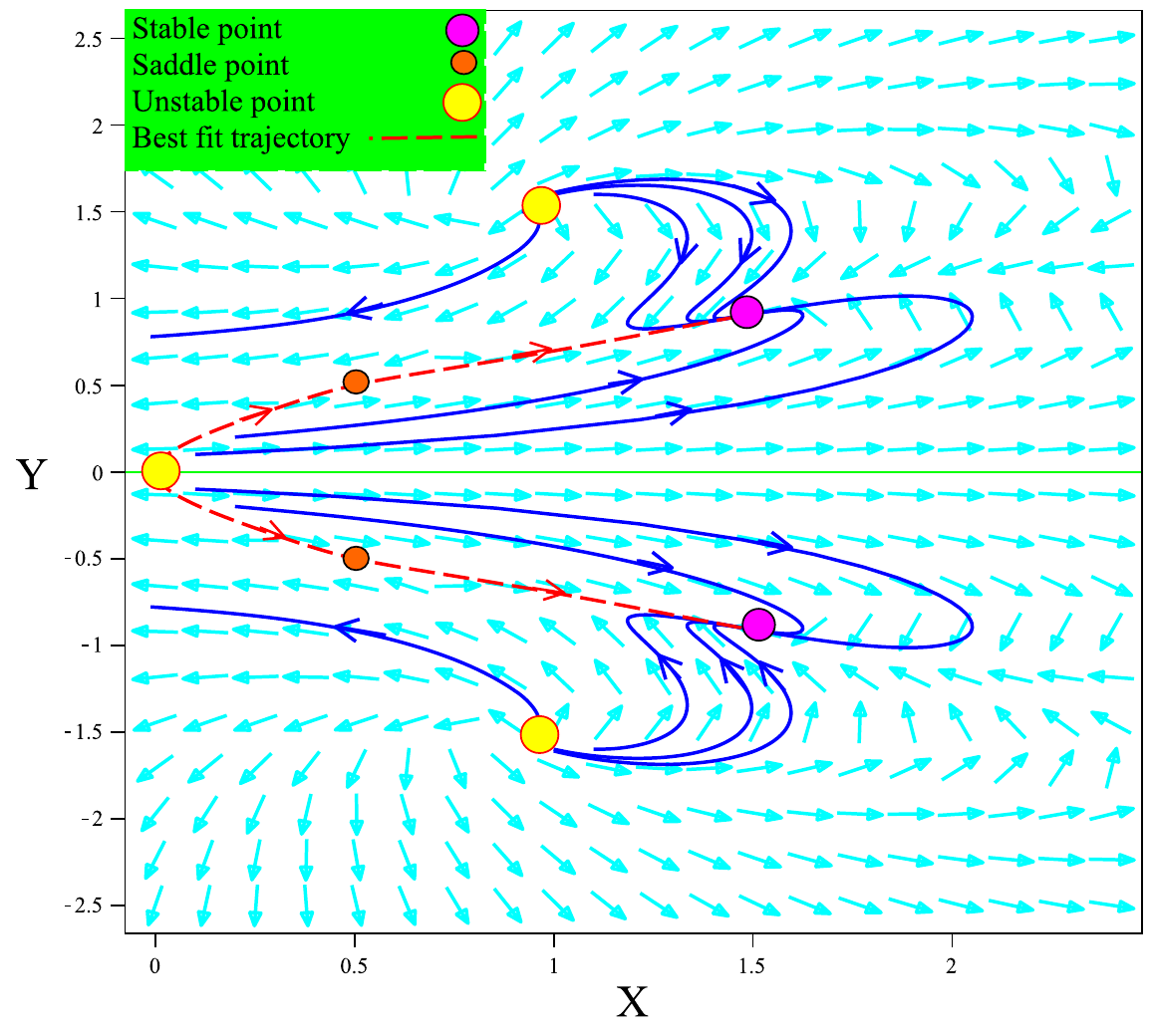}\hspace{0.1 cm}\\
Fig. 4: The attractor property of the dynamical system in
the phase plane.  The trajectories\\ with the model parameters $\alpha$ and $\beta$ are shown by blue lines.\\ The red trajectories correspond to the \\
best fitted stability parameter and initial conditions.
\end{tabular*}\\

In Fig. 5, in more details for the given initial condition of $Y > 0$, the trajectories (blue curves) approaching the stable critical point FP5 in the phase plane are shown. For the initial conditions of $Y <0$, then the trajectory shown entering the stable critical point FP6. The best fitted trajectories ( red curve) with the properties given in the previous section enter the stable critical points FP5 and FP6. These trajectories both fit the model parameters $\alpha$ and $\beta$ and the initial conditions for $X$, $Y$ and $H$.\\

\begin{tabular*}{2.5 cm}{cc}
\includegraphics[scale=.35]{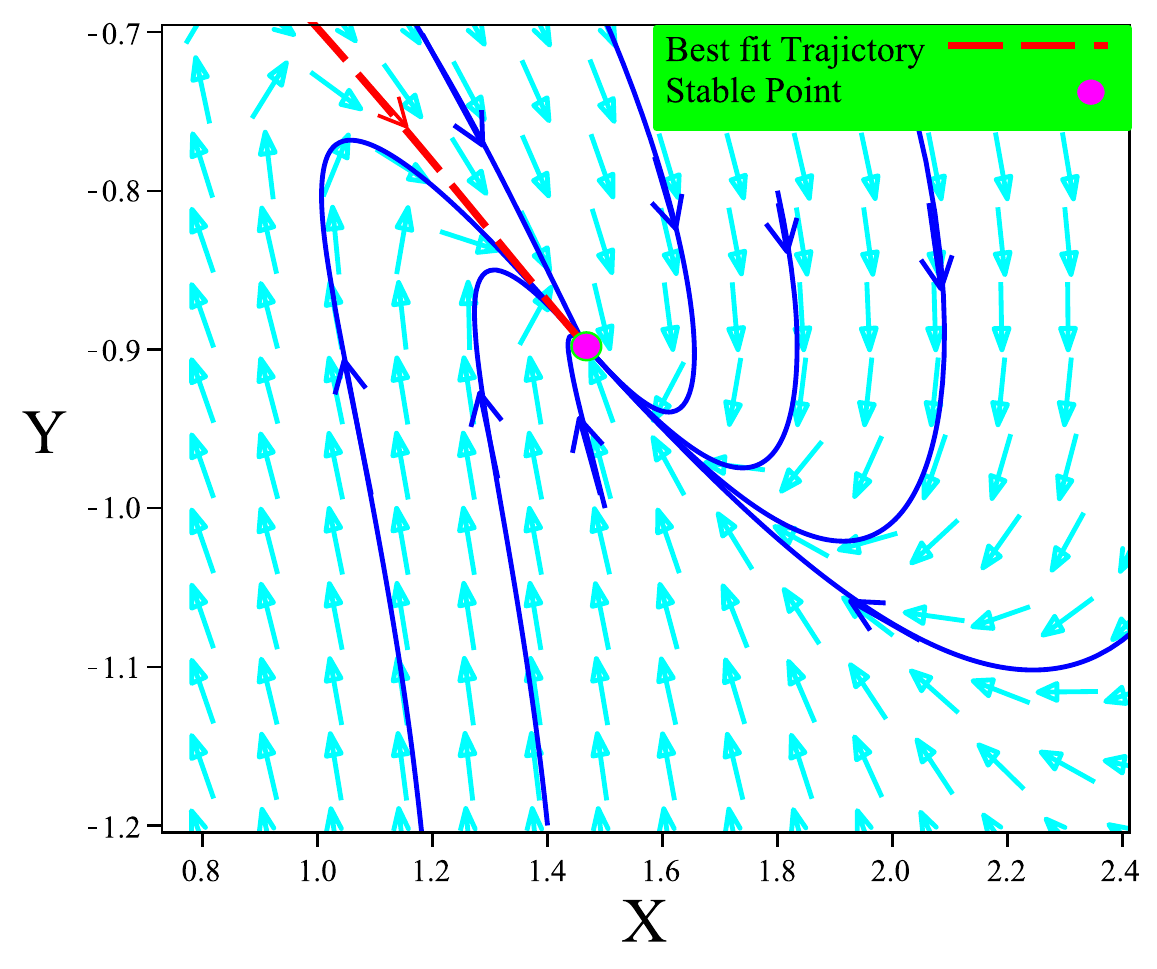}\hspace{0.1 cm}\includegraphics[scale=.35]{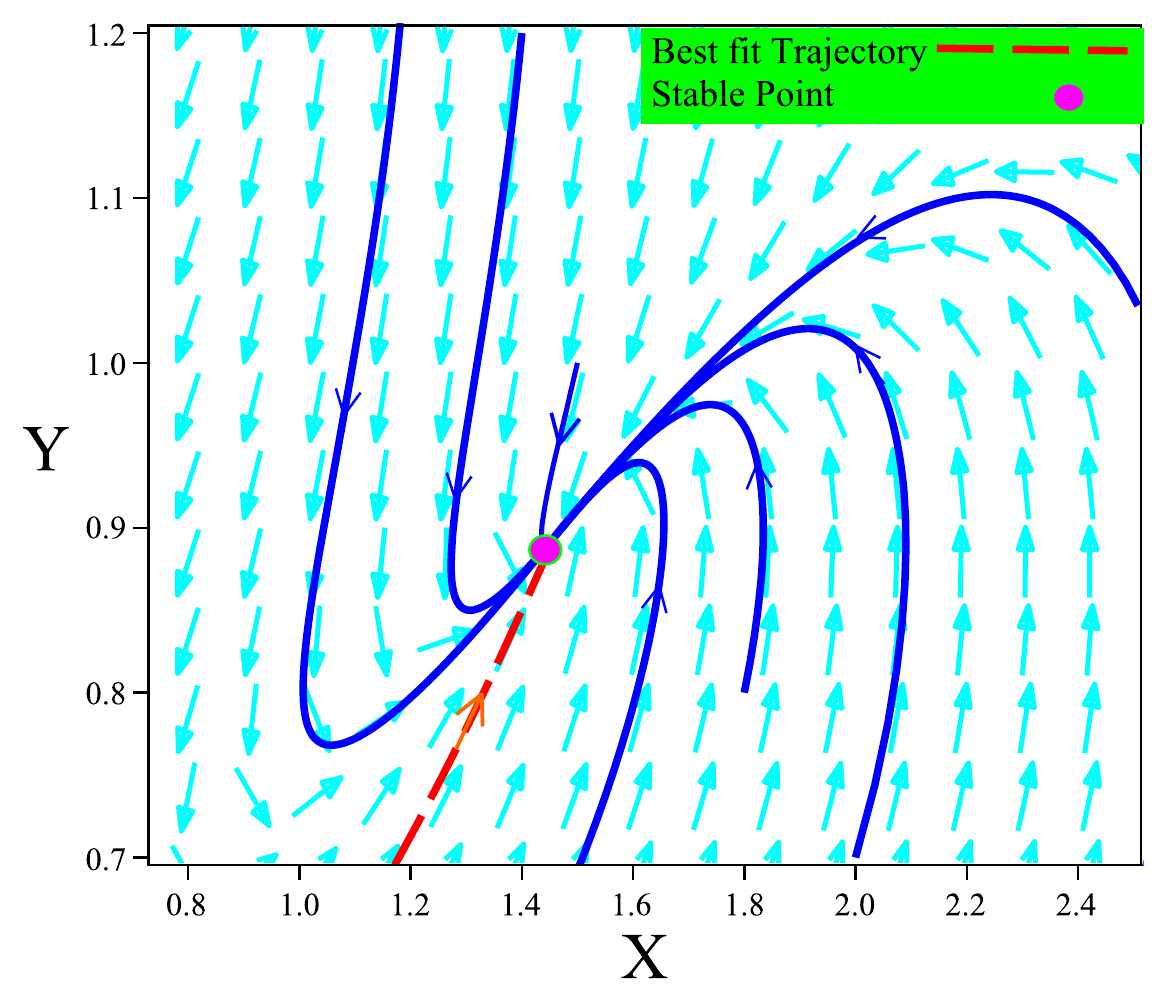}\hspace{0.1 cm}\\
Fig. 5: The attractor property of the dynamical system in
the phase plane.  The trajectories \\entering the best fitted stable critical points with the model parameters $\alpha$ and $\beta$ \\are shown by blue lines. The red trajectory approaching the FP5 and FP6 \\corresponds to the
best fitted stability parameter and initial conditions.
\end{tabular*}\\

\section{Cosmological parameters}

In order to understand the behavior of the universe and its dynamics we need to study the cosmological parameters such as EoS parameter. We have already verified our model with the current observational data via the distance modulus test. The EoS parameters analytically and/or numerically have been investigated by many authors for variety of cosmological models. Applying stability analysis and simultaneously best fitting the model with the observational data using $\chi^2$ method gives us a better understanding of the critical points. In our model, the effective EoS parameter is defined by $\omega_{eff}=-1-\frac{2}{3}\frac{\dot{H}}{H^{2}}$ where $\frac{\dot{H}}{H^{2}}$ is given in terms of the dynamical variables in equation (\ref{hdot}). Among cosmological parameters, the statefinders are given by $r=\ddot{H}/H^{3}-3q-2$ and $s=(r-1)/3(q-\frac{1}{2})$ discussed here, where $q$ in $r$ and $s$ is the deceleration parameter and $\frac{\ddot{H}}{H^{2}}$ in $r$ in terms of new dynamical variables can be obtained by taking derivative of $\dot{H}$.

The calculated effective EoS parameter for the best fitted stability parameters $\alpha$ and $\beta$ and initial conditions exhibits phantom crossing behavior in the future where the universe is in an unstable saddle point FP1 (Fig. 6). The current effective EoS parameter for the best fitted model parameters is $\omega_{eff} \simeq -0.667$. The graph also shows that in the far past for high redshift the universe is in matter dominated era in an unstable saddle point FP7 where $\omega_{eff}= 0$. The best fitted effective EoS parameter continues its journey towards unstable saddle point FP1 at about redshift $z\simeq -0.3$ where $\omega_{eff}= -1$ in near future. The interesting point is that the unstable saddle point state of the universe is when it crosses the phantom divide line. Finally, the trajectory approaches the stable critical point FP5 in the future.

\begin{tabular*}{2.5 cm}{cc}
\includegraphics[scale=.4]{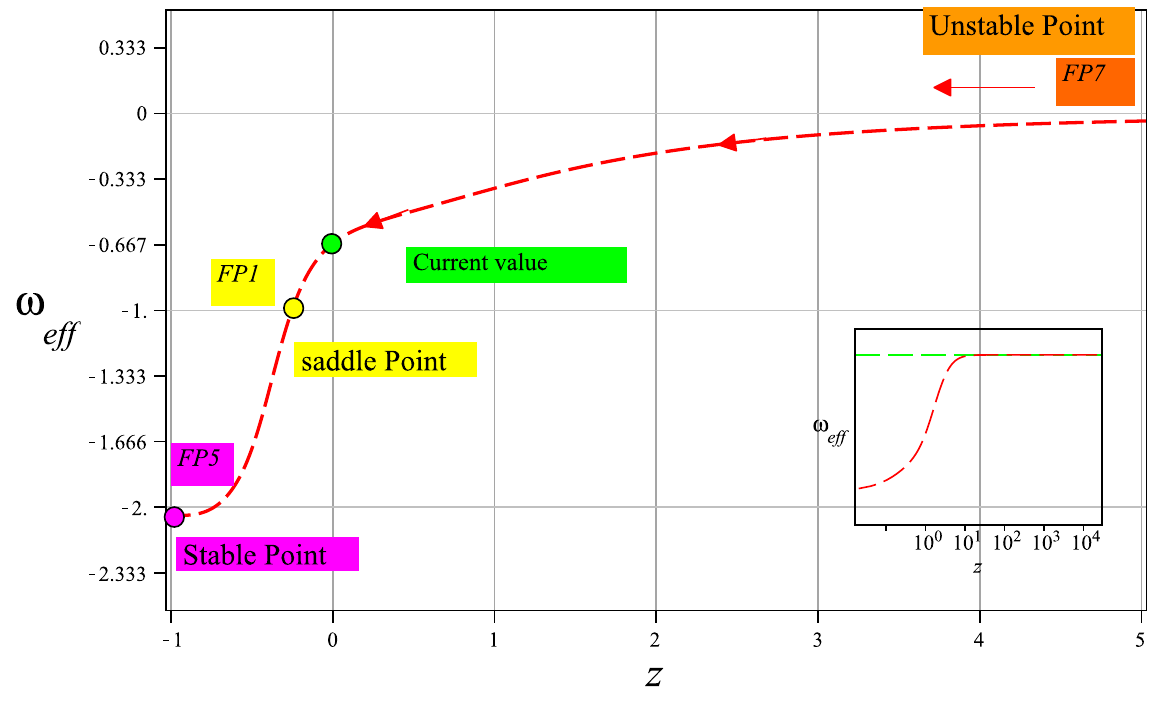}\hspace{0.1 cm}\\
Fig. 6: The best fit graph of effective $EoS$ parameter plotted as function of redshift
\end{tabular*}\\

Following \cite{cpl}, in Chevallier-Polarski-Linder (CPL) parametrization model one can use
linearly approximated EoS parameter,
\begin{equation}\label{cpl}
\omega_{CPL}=\omega_0+\omega_a\frac{z}{1+z},
 \end{equation}
where the parametrization is fitted for CPL parameters $\omega_0$ and $\omega_a$. In the following, using the $\chi^2$ method, we can best fit the CPL parameters with the observational data that have been used to find effective EoS parameter. Fig 7 shows the  $68.3\%$, $95.4\%$ and $99.7\%$ confidence level for $\omega_0$ and $\omega_a$ that lie within the green, blue and red contours, respectively.

\begin{tabular*}{2.5 cm}{cc}
\includegraphics[scale=.35]{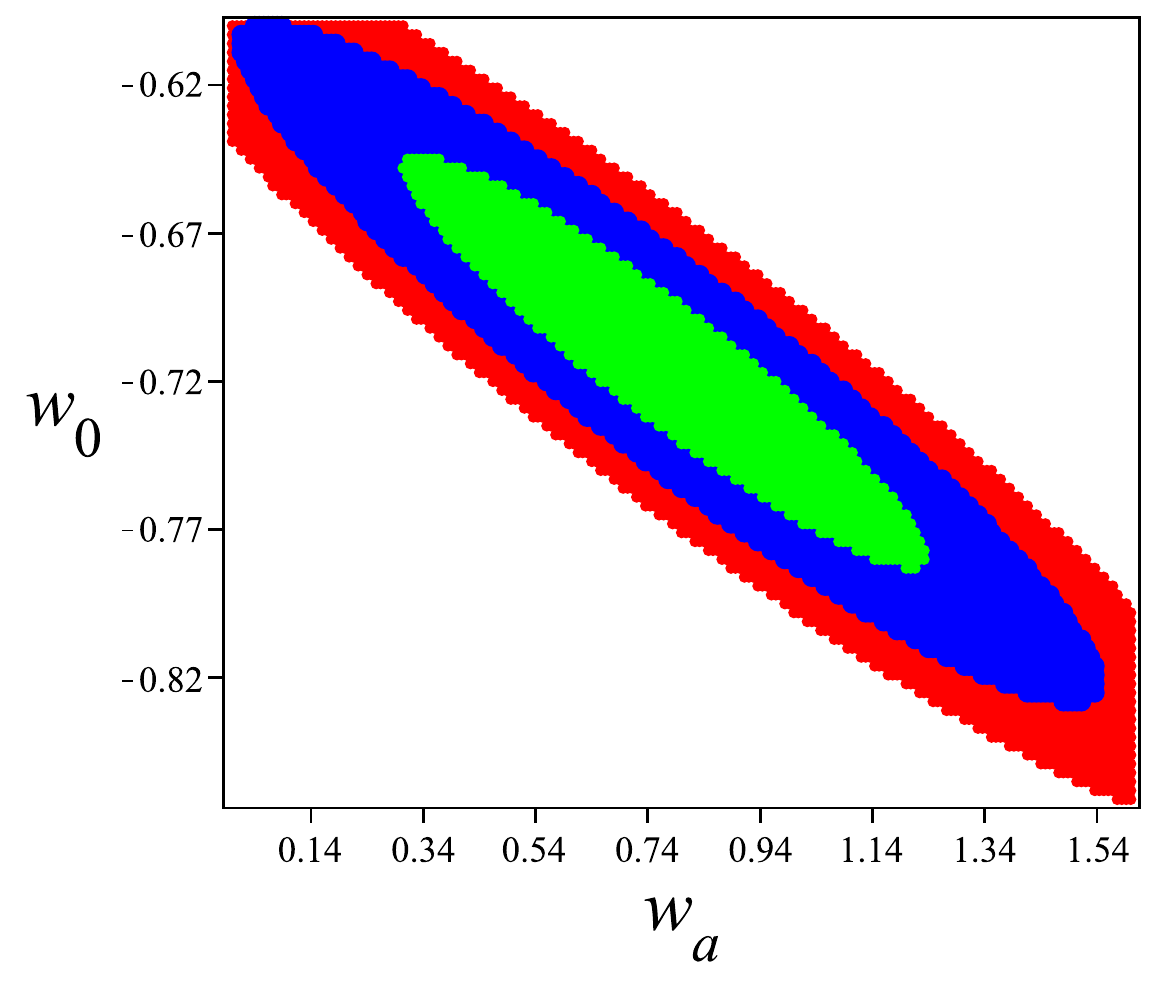}\hspace{0.1 cm}\\
Fig. 7: The constraint on $\omega_0$ and $\omega_a$ at the 68.3\%, 95.4\% and 99.7\% confidence levels.
\end{tabular*}\\

Alternatively, the 1-dim and 2-dim likelihood for the CPL parameters $\omega_0$ and $\omega_a$ are plotted in Fig. 8.

\begin{tabular*}{2.5 cm}{cc}
\includegraphics[scale=.25]{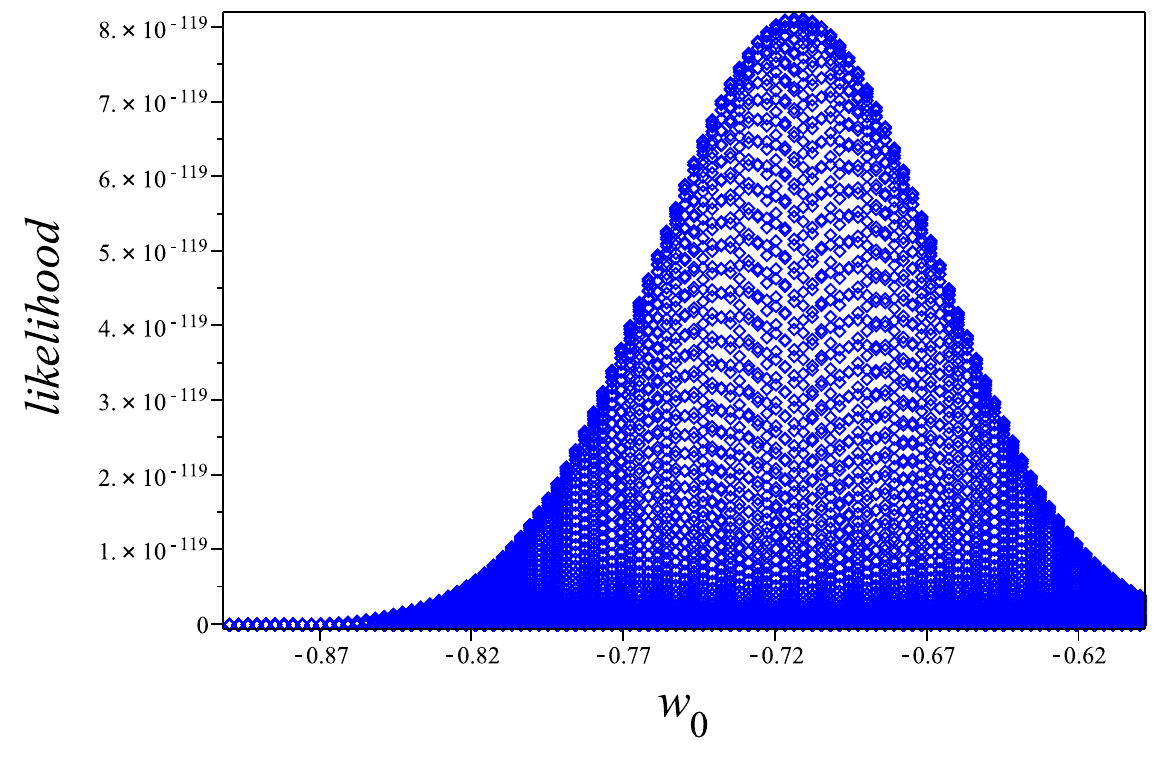}\hspace{0.1 cm}\includegraphics[scale=.25]{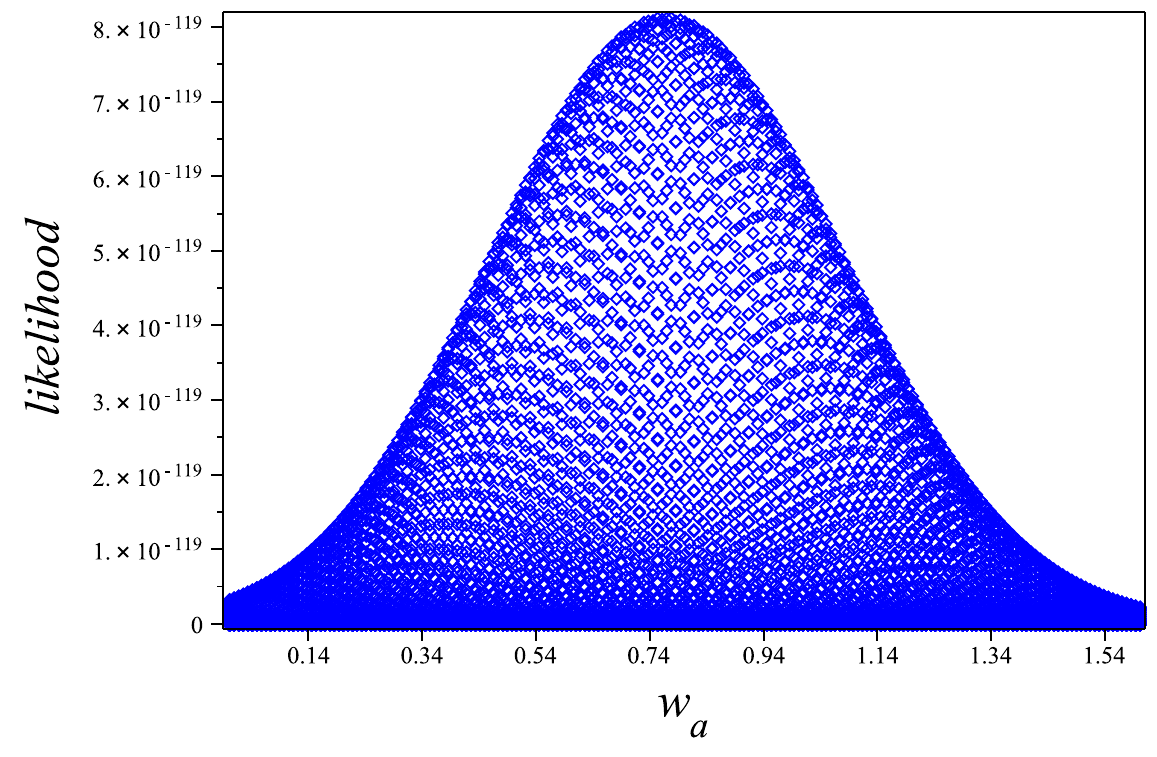}\hspace{0.1 cm}\includegraphics[scale=.3]{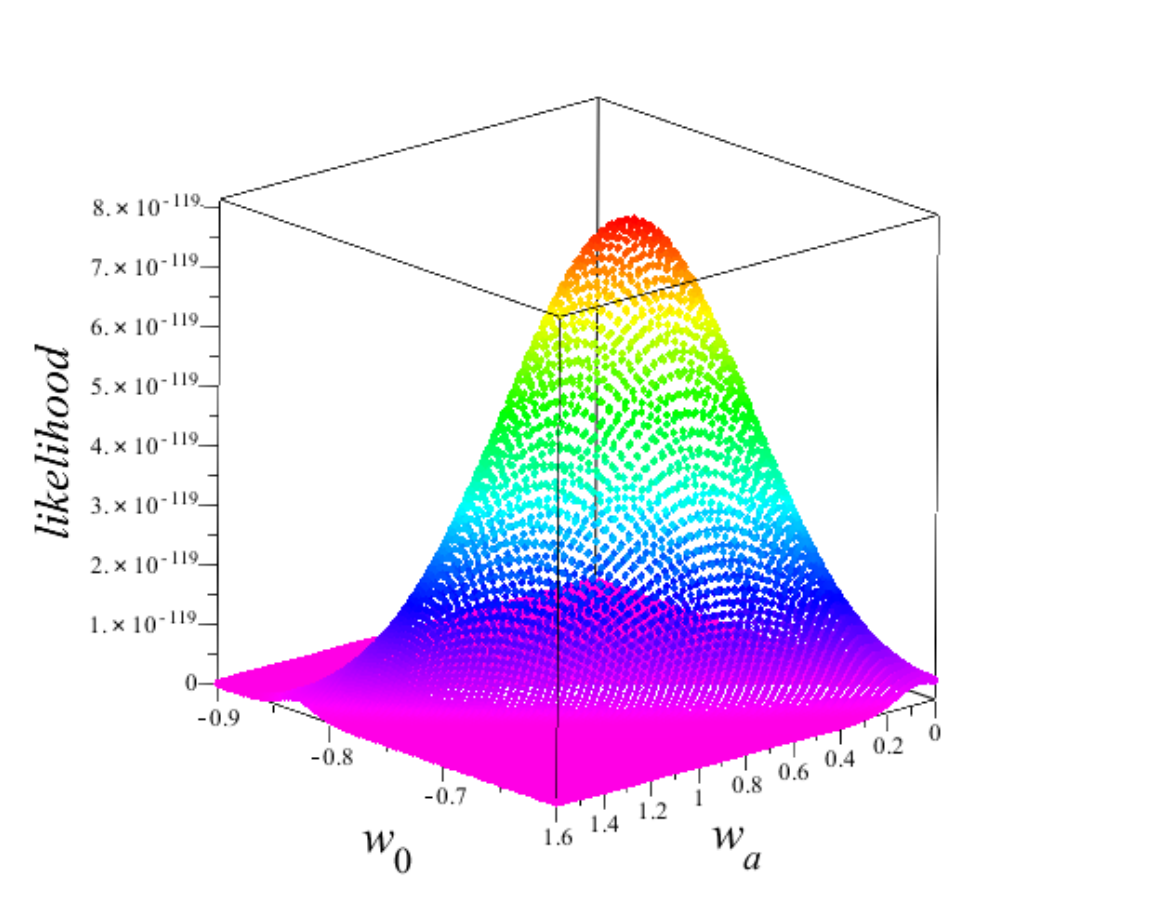}\hspace{0.1 cm}\\\\
Fig. 8: 1-dim and 2-dim likelihood for the CPL model parameters $\omega_0$ and $\omega_a$.
\end{tabular*}\\

The best fitting procedure thus finds that the approximate values of the CPL parameters are $\omega_0=-0.72$, $\omega_a=0.74$. A comparison of the standard CPL parametrisation matching the best fitted effective EoS parameter is shown in Fig. 9.

\begin{tabular*}{2.5 cm}{cc}
\includegraphics[scale=.4]{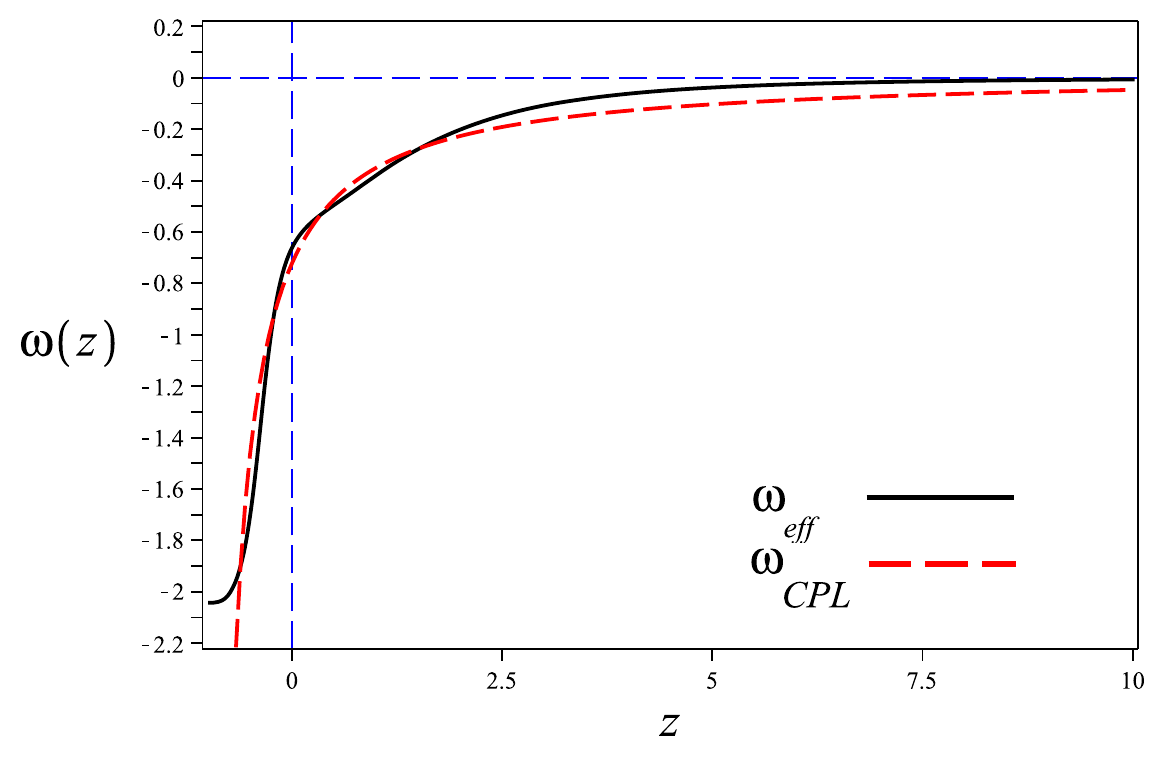}\hspace{0.1 cm}\\
Fig. 9: The best fitted effective $EoS$ parameter in comparison with the CPL EoS \\parameterization with the best approximated parameters  $\omega_0=-0.72$, $\omega_a=0.74$
\end{tabular*}\\

By applying the statefinder diagnosis to the model, Fig 10 shows the best-fitted trajectories of the statefinder diagrams $\{r,q\}$, $\{s,q\}$ and $\{r,s\}$. From the graph it can be seen that the best-fitted trajectory passes LCDM state with $\{r, s\}=\{1, 0\}$ sometimes in the past.
The current value of the best fitted trajectory and its location with respect to
the LCDM state can also be observed in the $\{r,s\}$ diagram. From Fig 10, we see that the current value of the statefinder $\{r, s\}$ is close to the LCDM state and compatible with the recent observational data.

\begin{tabular*}{2.5 cm}{cc}
\includegraphics[scale=.25]{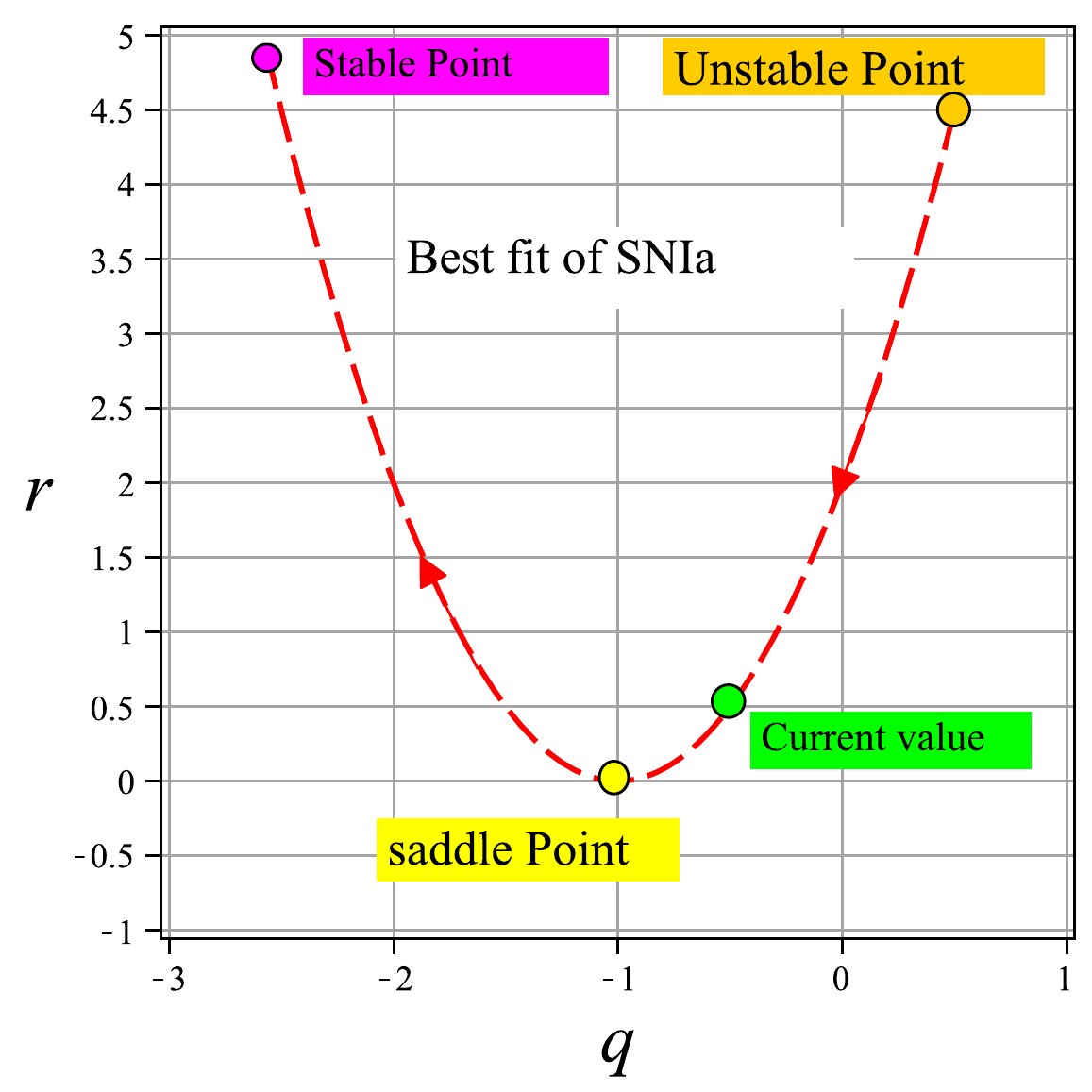}\hspace{0.1 cm}\includegraphics[scale=.25]{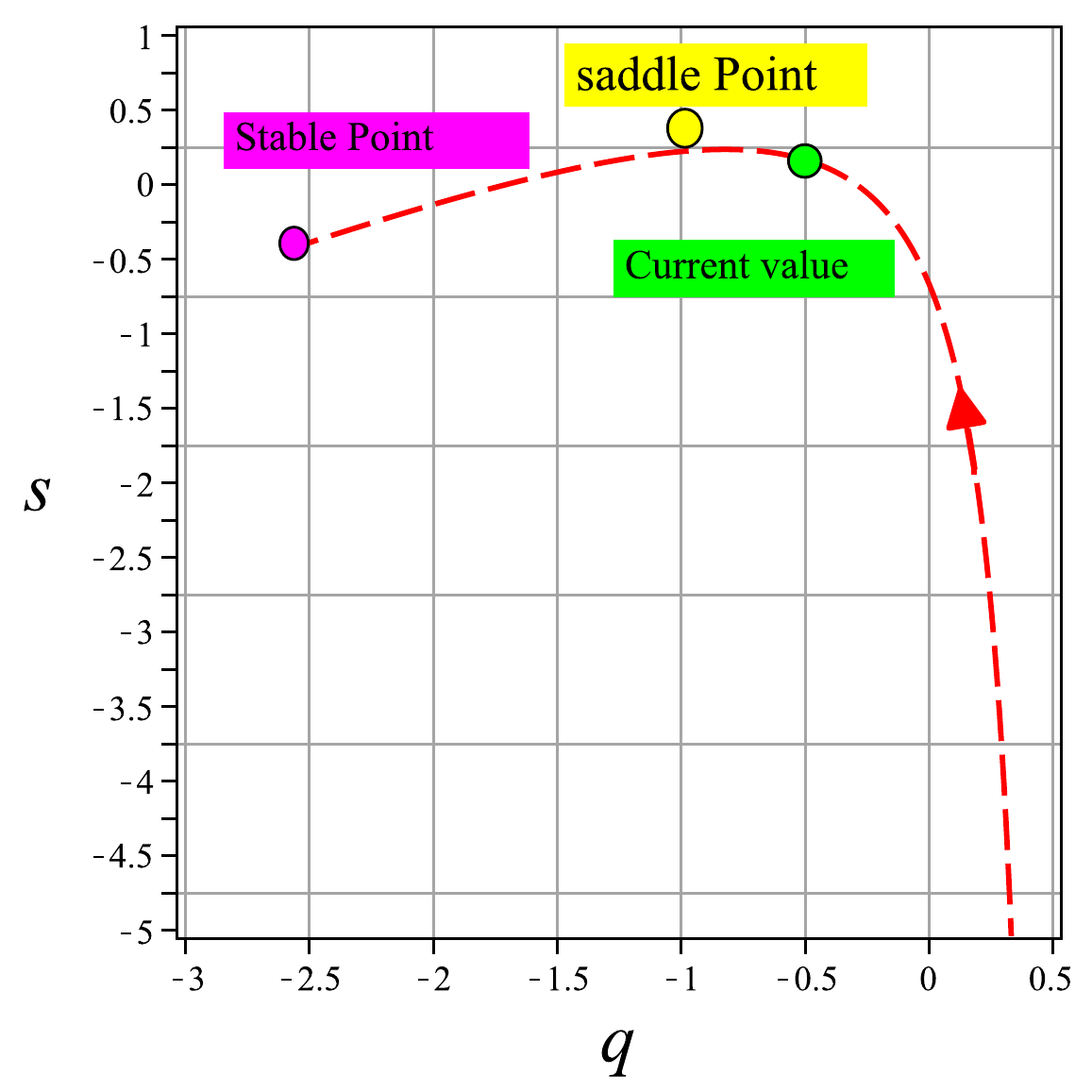}\hspace{0.1 cm}\includegraphics[scale=.25]{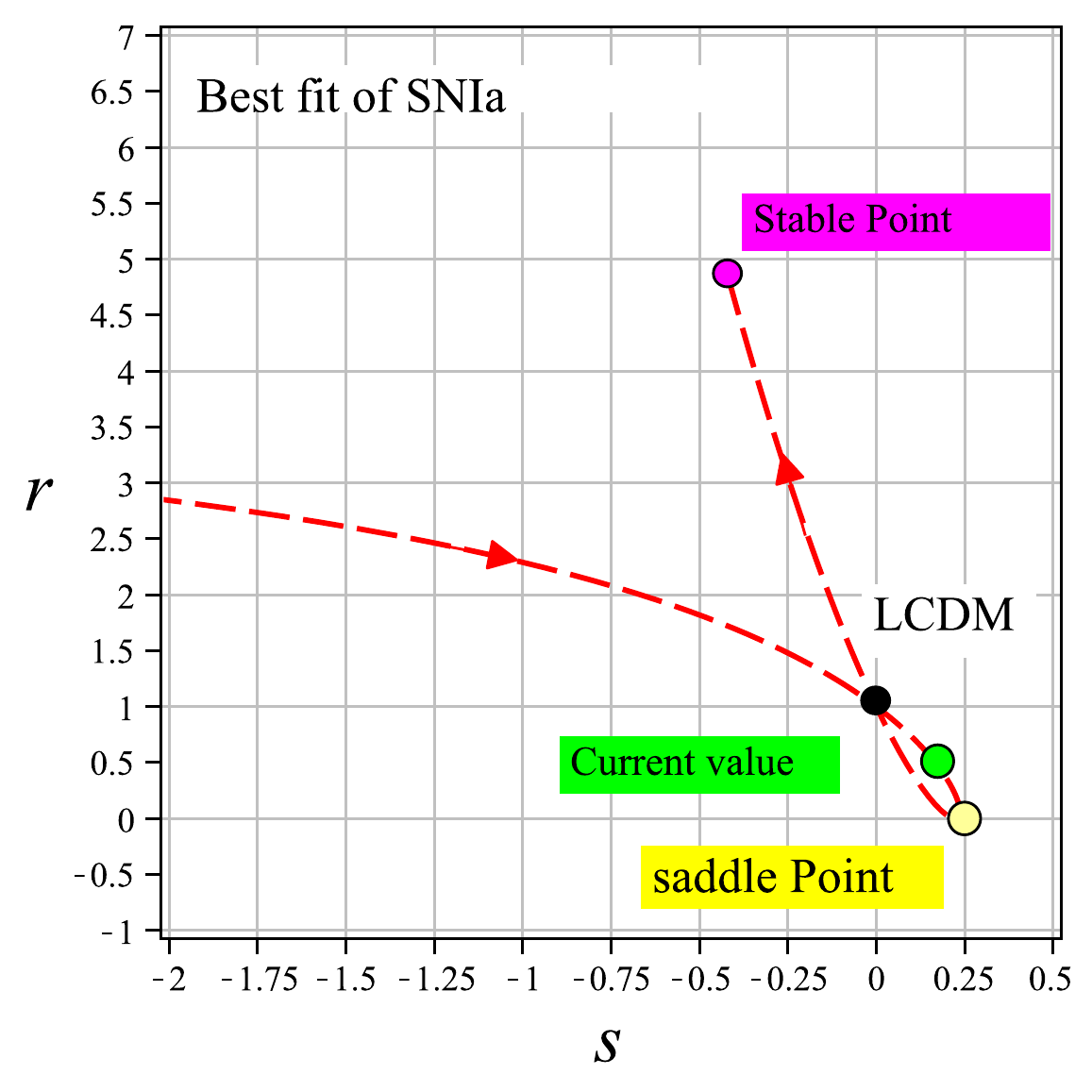}\hspace{0.1 cm}\\
Fig. 10: The best fit graph of statefinders $\{r,q\}$, $\{s,q\}$ and $\{r,s\}$
\end{tabular*}\\

In Fig 11 we depict the corresponding dynamical behavior of the satefinder  $\{r,s\}$ against $N=ln( a)$. From Figs. 10 and 11 we observe that the universe starts its journey from unstable state in the past, passed the current state and eventually reaches a stable state in the future. Interestingly, the trajectories of the statefinders and their evolution show that sometimes in the future the universe approaches an unstable saddle point that is the extreme points of the statefinder trajectories corresponding to the state that the universe crosses phantom divide line.

\begin{tabular*}{2.5 cm}{cc}
\includegraphics[scale=.25]{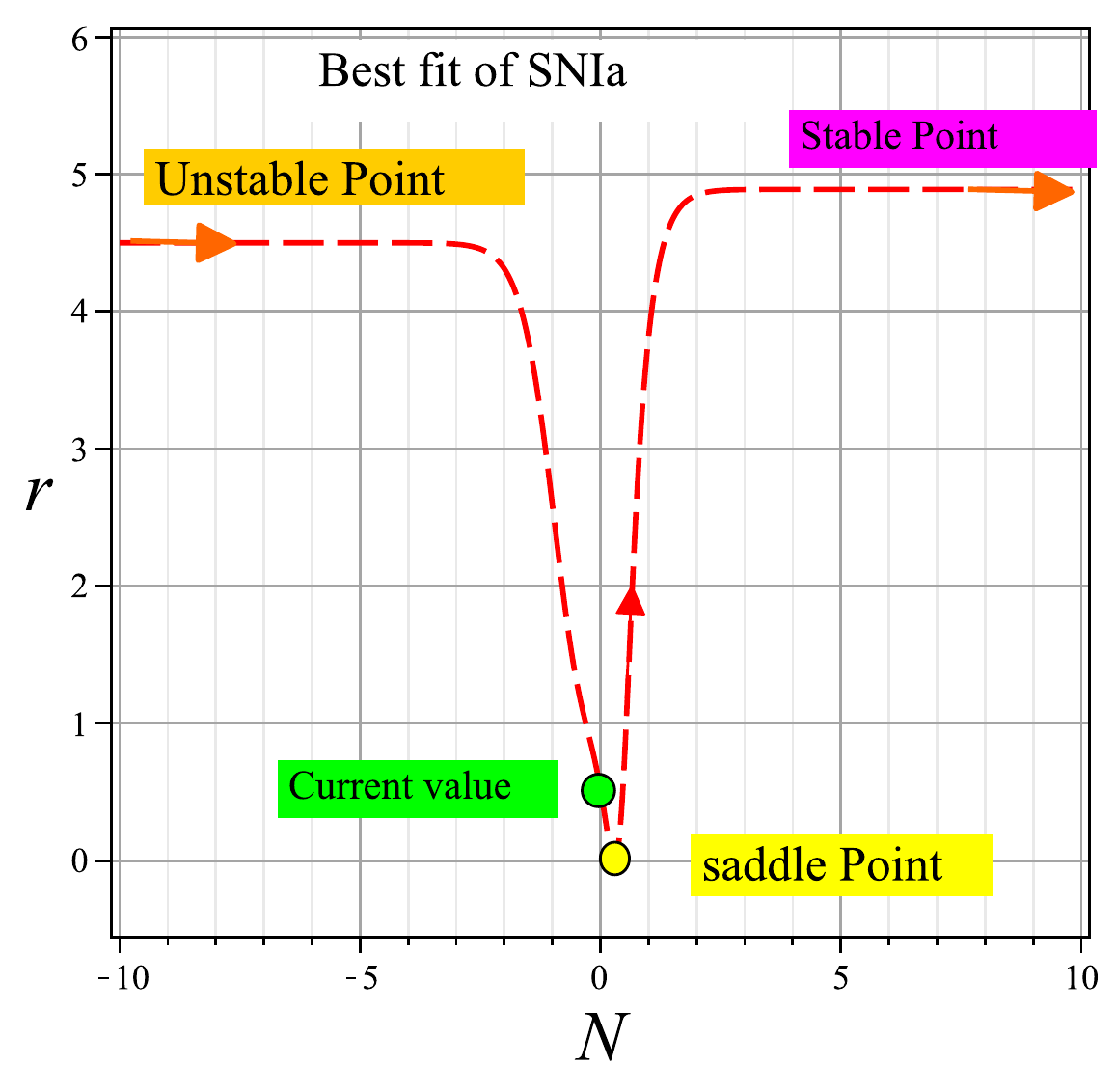}\hspace{0.1 cm}\includegraphics[scale=.25]{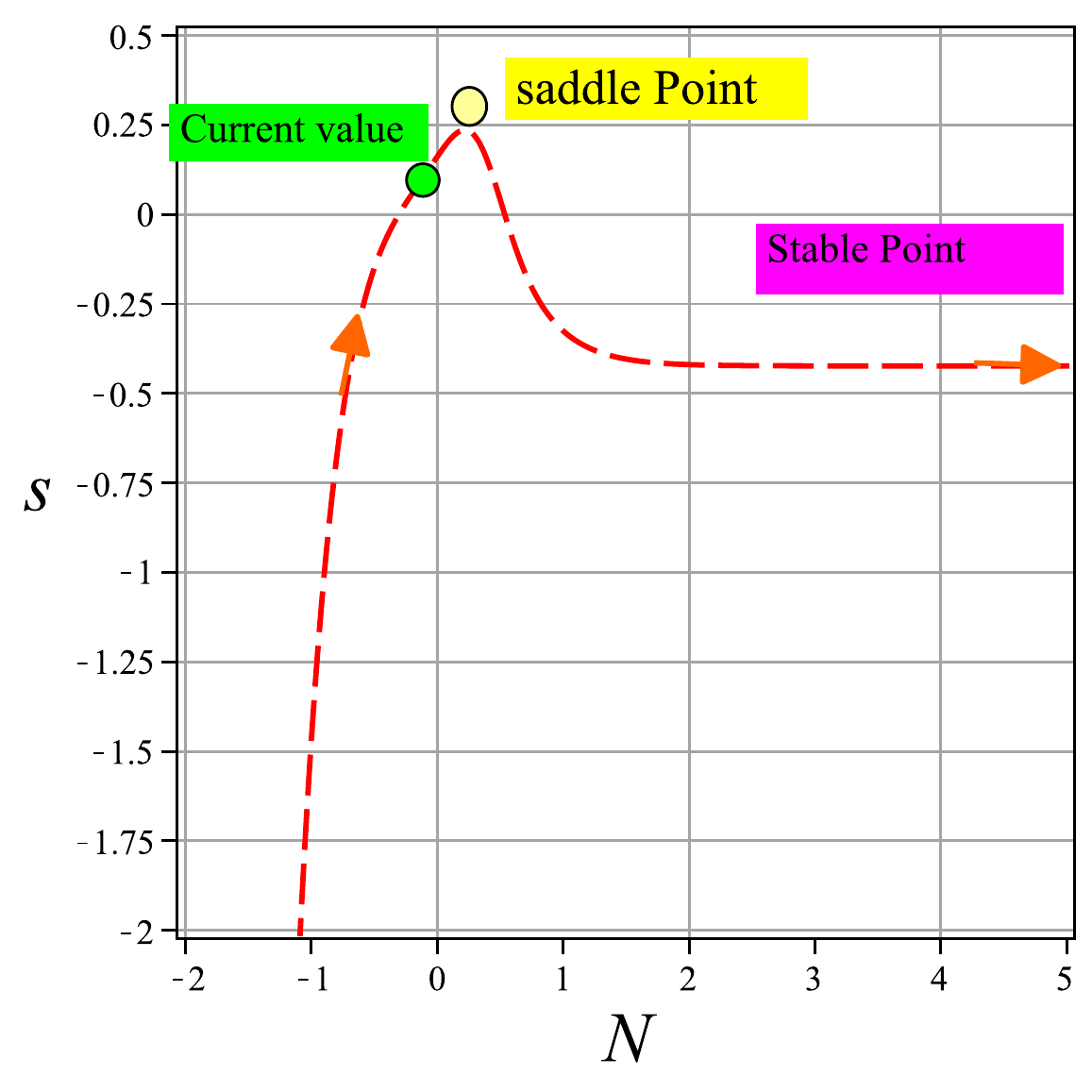}\hspace{0.1 cm}\\
Fig. 11: The best fit graph of statefinder parameters $r$ and $s$ plotted as function of $N=ln (a)$
\end{tabular*}\\

We also reconstructed the best-fitted functions $G(\phi)$ and $f(\phi)$. From Fig 12, we see that the trajectories for the best fitted model parameters illustrate a monotonic increasing and decreasing behavior with respect to the scalar field $\phi$ for $f(\phi)$ and $G(\phi)$ respectively.

\begin{tabular*}{2.5 cm}{cc}
\includegraphics[scale=.3]{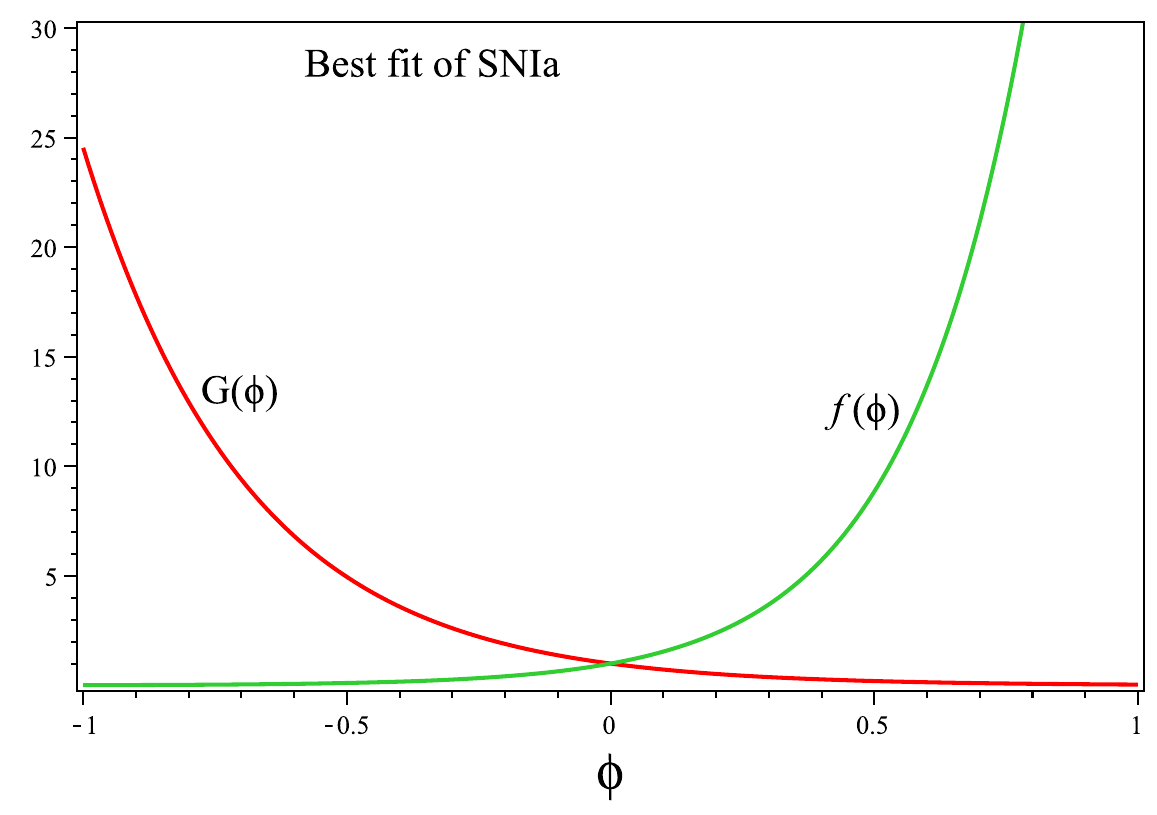}\hspace{0.1 cm}\\
Fig. 12: The best fit reconstructed graph of $G(\phi)$and $f(\phi)$ plotted as function of $\phi$
\end{tabular*}\\

\section{summary and remarks}

This paper is designed to study dynamics of the nonminimally coupled scalar field theory by
using the stability analysis and the 2-dimensional phase space of the theory. In a different approach in stability analysis, here, we solve the system of autonomous differential equations by best fitting the model parameters and also the initial conditions with the observational data for distance modulus. Therefore all the critical points with the stability conditions are presented in the model are physically reliable and observationally verified. By stability analysis, we find two stable critical points in the model as shown in Fig. 4 and 5. With the best fitting results the trajectories approaching the stable points are observationally verified.

We then study the cosmological parameters such as effective EoS parameter, $\omega_{eff}$ for the model in terms of the dynamical variables introduced in the stability section. The result show that with best fitted stability parameter, the model exhibits an accelerating universe with a transition from $\omega_{eff}>-1/3$ (decelerating universe) to $\omega_{eff}<-1/3$ (accelerating universe)in the past. It also shows that the universe sometimes in the future passes an unstable state ( saddle critical point in the phase space) where the effective EoS parameter crosses the phantom divide line. The best fitted effective EoS parameter is also compared with the CPL parametrisation of EoS and the best approximate value of the CPL parameters found to be $\omega_0=-0.72$, $\omega_a=0.74$. With the statefinder diagnostic, the best fitted statefinder parameters show that the current state of the universe is very close to LCDM. The statefinder graphs and also their evolution explicitly illustrate that the unstable saddle point is the extreme point of the statefinder trajectories. Finally, the best fitted reconstructed coupled scalar field functions, $\dot{\phi}$, $G(\phi)$ and $f(\phi)$ against the scalar field display a smoothly decreasing and increasing behavior.

\section{Acknowledgement}
The authors would like to thank the anonymous reviewer for generous comments that led to considerable improvement of this paper.
We would also like to thank University of Guilan Research Council for its support.


\begin{thebibliography}{99}


\bibitem{Reiss} A.G. Reiss et al,[Supernova Search TeamCollaboration] Astron. J. 116, 1009 (1998); S. Perlmutter et al, Astrophys
J. 517 565(1999) ; J. L. Tonry et al, Astrophys J. 594, 1-24 (2003)

\bibitem{Bennet} C. I. Bennet et al, Astrophys J. Suppl. 148:1, (2003); C. B.
 Netterfield et al, Astrophys J. 571, 604 (2002); N. W. Halverson et al, Astrophys J. 568, 38 (2002)

\bibitem{Pope} A. C. Pope, et. al, Astrophys J. 607 655, (2004)

\bibitem{Riess2}A.G. Riess et al., Astrophys. J. 607 (2004) 665; R. A. Knop et al, Astrophys. J. 598 (2003) 102.

\bibitem{Abazajian}K. Abazajian et al, Astron. J. 129 (2005) 1755; Astron. J. 128 (2004) 502; Astron. J. 126 (2003) 2081; M. Tegmark et al,
Astrophys. J. 606 702 (2004)

\bibitem{Allen}S.W. Allen, R. W. Schmidt, H. Ebeling, A. C. Fabian and L. van Speybroeck, Mon. Not. Roy. Astron. Soc. 353 457 (2004)

\bibitem{Bennett}C.L. Bennett et al, Astrophys. J. Suppl. 148 1 (2003)

\bibitem{Spergel} D. N. Spergel, et. al., Astrophys J. Supp. 148 175, (2003)

\bibitem{Seljak}U. Seljak et al, Phys. Rev. D 71 (2005) 103515; M. Tegmark, JCAP 0504 001(2005)

\bibitem{Boisseau} B. Boisseau, G. Esposito-Farese, D. Polarski, A.A. Starobinsky, 	Phys.Rev.Lett. 85 2236 (2000)

\bibitem{Setare} M. R. Setare, Phys. Lett. B644:99-103,(2007); J. Sadeghi, M. R. Setare, A. Banijamali,  Eur. Phys. J. C64:433-438,(2009); J. Sadeghi, M. R. Setare,
  A. Banijamali, Phys. Lett. B679:302-305,(2009); J. Sadeghi, M. R. Setare,  A. Banijamali,  Phys. Lett. B678:164-167,(2009)


\bibitem{Amendola} L. Amendola, D. T. Valentini; Phys.Rev. D64 043509 (2001); L. Amendola, R. Gannouji, D. Polarski, S. Tsujikawa; Phys.Rev. D75 083504 (2007)


\bibitem{capelo}S. Capozziello, G. Lambiase, Grav.Cosmol. 6, 164-172 (2000)


\bibitem{Sahoo}B.K. Sahoo and L.P. Singh, Mod. Phys. Lett. A 17 2409(2002); B.K. Sahoo and L.P. Singh, Mod. Phys. Lett. A 18 2725 (2003)
; B.K. Sahoo and L.P. Singh, Mod. Phys. Lett. A 19 1745(2004); J. Sadeghi, M.R. Setare, A. Banijamali and F. Milani, Phys. Rev. D 79 123003(2009)

\bibitem{Capozziello}S. Capozziello, S. Carloni and A. Troisi, Recent Res. Dev. Astron.
Astrophys. 1 (2003) 625; S. Nojiri and S.D. Odintsov, Phys. Rev. D 68 (2003) 123512; Phys. Lett. B 576 5(2003)

\bibitem{Faraoni} V. Faraoni, Phys. Rev. D 75 067302(2007); J.C.C. de Souza, V. Faraoni, Class. Quant. Grav. 24 3637(2007); A.W. Brookfield, C.
van de Bruck and L.M.H. Hall, Phys. Rev. D 74 064028 (2006); F. Briscese, E. Elizalde, S. Nojiri and S.D. Odintsov, Phys. Lett. B
646 105 (2007); S. Rahvar and Y. Sobouti, Mod. Phys. Lett. A 23 1929 (2008); O. Bertolami, C.G. Bohmer, T. Harko and F.S.N. Lobo,
Phys. Rev. D 75 104016 (2007); S. Carloni, P.K.S. Dunsby and A. Troisi, Phys. Rev. D 77 024024 (2008); F. Briscese and E.
Elizalde, Phys. Rev. D 77 044009 (2008); B. Li and J.D. Barrow, Phys. Rev. D 75 084010 (2007)

\bibitem{Nojiri3}S. Nojiri and S.D. Odintsov, Gen. Rel. Grav. 36 1765 (2004); Phys. Lett. B 599 137 (2004); G. Cognola, E. Elizalde, S. Nojiri, S.D.
Odintsov and S. Zerbini, JCAP 0502 010 (2005); Phys. Rev. D 73 084007 (2006); K. Henttunen, T. Multamaki and I. Vilja, Phys. Rev.
D 77 024040 (2008); T. Clifton and J. D. Barrow, Phys. Rev. D 72 103005 (2005); T. Koivisto, Phys. Rev. D 76 043527 (2007); S.K.
Srivastava, Phys. Lett. B 648 119 (2007); S. Nojiri, S.D. Odintsov and P. Tretyakov, Phys. Lett. B 651 224 (2007); S. Baghram, M.
Farhang and S. Rahvar, Phys. Rev. D 75 044024 (2007); H. Farajollahi, F. Milani, Mod. Phys. Lett. A 25:2349-2362 (2010)

\bibitem{Setare1} M. R. Setare, M. Jamil, Phys. Lett. B 690  1-4 (2010); A. C. Davis, C. A.O. Schelpe, D. J. Shaw, Phys.Rev.D80:064016 (92009);
Y. Ito, S. Nojiri, Phys.Rev.D79:103008 (2009); Takashi Tamaki, Shinji Tsujikawa, Phys.Rev.D78:084028 (2008)

\bibitem{Mota1} D.F. Mota, D.J. Shaw, Phys. Rev. D 75 063501 (2007); K. Dimopoulos, M. Axenides, JCAP 0506:008
(2005); T. Damour, G. W. Gibbons and C. Gundlach, Phys. Rev. Lett, 64, 123 (1990); H. Farajollahi, N. Mohamadi, Int.J.Theor.Phys.49:72-78 (2010);
H. Farajollahi, N. Mohamadi, H. Amiri, Mod. Phys. Lett. A, 25, No. 30 2579-2589 (2010)

\bibitem{Carr} S. M. Carroll, Phys. Rev. Lett. 81 3067(1998); S. M. Carroll, W. H. Press and E. L. Turner, Ann. Rev. Astron. Astrophys, 30, 499 (1992); T. Biswas and A. Mazumdar, arXiv:hep-th/0408026; T. Biswas, R. Brandenberger, A. Mazumdar and T. Multamaki. Phys.Rev. D74 , 063501(2006)

\bibitem{Amendola1}L. Amendola; Phys.Rev. D60 043501(1999); R. Gannouji, D. Polarski, A. Ranquet, A. Starobinsky; JCAP 0609 016(2006); L. Perivolaropoulos; JCAP 0510  001(2005); N. Bartolo, M. Pietroni Scalar tensor gravity and quintessence Phys.Rev.
D61 023518 (2000); F. Perrotta, C. Baccigalupi, S.Matarrese; Phys.Rev. D61 023507(2000)

\bibitem{Sahni} V. Sahni, T. D. Saini, A. A. Starobinsky, U. Alam, JETPLett.77:201-206 (2003)

\bibitem{Alam} U. Alam, V. Sahni, T. D. Saini and A. A. Starobinsky, Mon. Not. Roy. Astron. Soc. 344, 1057 (2003)

\bibitem{Zimdahl} W. Zimdahl, D. Pavon, Gen. Rel. Grav. 36, 1483 (2004); X. Zhang, Phys. Lett. B611, 1 (2005)

\bibitem{Yi} Z. L. Yi and T. J. Zhang, Phys. Rev. D75, 083515 (2007)

\bibitem{faraj2} H. Farajollahi, A. Salehi, Phys.Rev.D83:124042, (2011); H. Farajollahi, A. Salehi, F. Tayebi, A. Ravanpak, JCAP 05 017 (2011); Hossein Farajollahi, Amin Salehi, JCAP 1011:006,(2010); H. Farajollahi, A. Salehi, F. Tayebi, Astrophys Space Sci 335:629-634(2011)

\bibitem{Linder} E.V. Linder, Rep. Prog. Phys. 71 056901 (2008)

\bibitem{stability} H. Farajollahi, A. Salehi, JCAP 1011:006 (2010)

\bibitem{Chiang} Chiang-Mei Chen, W.F. Kao, Phys.Rev. D64 124019 (2001); T. C. Charters, A. Nunes, J. P. Mimoso, Class. Quant. Grav.18:1703-1714 (2001); V. V. Kiselev, JCAP0801:019 (2008); H. Farajollahi, F. Milani, Int. J. Theor. Phys. 50, 6, 1953-1961 (2011)


\bibitem{Birrel} N.D. Birrel and P.C.W. Davies, Quantum Field in Curved Space (Cambridge Univ. Press, Cambridge, 1986).


\bibitem{mota} David F. Mota, Douglas J. Shaw, 	Phys. Rev. D 75:063501,(2007)

\bibitem{cpl} M. Chevallier and D. Polarski, Int. J. Mod. Phys. D 10, 213 (2001); E. V. Linder, Phys. Rev. Lett. 90, 091301 (2003)


\end{thebibliography}
\end{document}